\documentclass[sigconf,nonacm]{acmart}

\AtBeginDocument{%
  }

\acmYear{2025}
\setcopyright{cc}
\setcctype{by}
\acmConference[CHI '25]{CHI Conference on Human Factors in Computing Systems}{April 26-May 1, 2025}{Yokohama, Japan}
\acmBooktitle{CHI Conference on Human Factors in Computing Systems (CHI '25), April 26-May 1, 2025, Yokohama, Japan}

\begin{document}

\title[Evaluating Eye Tracking and Electroencephalography as Indicator for Selective Exposure]{Evaluating Eye Tracking and Electroencephalography as Indicator for Selective Exposure During Online News Reading}

\author{Thomas Krämer}
\email{thomas.kraemer@gesis.org}
\orcid{0000-0003-0507-7843}
\affiliation{
  \institution{GESIS Leibniz Institute for the Social Sciences}
  \city{Cologne}
  \state{DE-NW}
  \country{Germany}
}

\author{Francesco Chiossi}
 \orcid{0000-0003-2987-7634}
\affiliation{
  \institution{LMU Munich}
  \city{Munich}
  \country{Germany}
}
\email{francesco.chiossi@lmu.de}

\author{Thomas Kosch}
\orcid{0000-0001-6300-9035}
\affiliation{
  \institution{HU Berlin}
  \city{Berlin}
  \country{Germany}}
\email{thomas.kosch@hu-berlin.de}

\renewcommand{\shortauthors}{Krämer et al.}

\begin{abstract}

Selective exposure to online news consumption reinforces filter bubbles, restricting access to diverse viewpoints. Interactive systems can counteract this bias by suggesting alternative perspectives, but they require real-time indicators to identify selective exposure. This workshop paper proposes the integration of physiological sensing, including Electroencephalography (EEG) and eye tracking, to measure selective exposure. We propose methods for examining news agreement and its relationship to theta band power in the parietal region, indicating a potential link between cortical activity and selective exposure. Our vision is interactive systems that detect selective exposure and provide alternative views in real time. We suggest that future news interfaces incorporate physiological signals to promote more balanced information consumption. This work joins the discussion on AI-enhanced methodology for bias detection.

\end{abstract}




\maketitle
\section{Introduction and Background}
Humans form mental models when interacting with online news based on what they believe, prefer and are familiar with~\cite{CARROLL198845}. In this context, selective exposure describes the tendency to consume online news that aligns with the worldviews of the user while avoiding contradictory information~\cite{boonprakong4797710scopingreview, boonprakong2025assessing}. Understanding selective exposure in Human-Computer Interaction (HCI) can inform interactive systems, for example, bias-aware systems, that promote diverse viewpoints~\cite{Dingler2023}. 

Prior research has examined selective exposure through behavioral metrics, including gaze duration and fixation patterns \cite{schmuck2020avoiding}. However, fixations alone do not reliably indicate cognitive engagement \cite{foulsham2013mind}. Psychophysiological studies suggest theta band activity as a marker for memory encoding and decision-consistent processing \cite{Fischer2013, Klimesch2001, chiossi2025designing, long2024multimodal}, but these studies often rely on static stimuli rather than self-selected reading contexts~\cite{boonprakong2025assessing}.

To address these gaps, this workshop paper suggests a follow-up study of our previous publication~\cite{kraemer2025escaping}, where we analyzed the feasibility of Electroencephalography (EEG) and eye tracking for quantifying selective exposure in an online news reading environment. Unlike prior unimodal approaches using static stimuli \cite{Fischer2013, Klimesch2001, schmuck2020avoiding}, we analyze gaze and EEG measures while participants read online news~\cite{schneegassEtAl2020BrainCoDe, KoschEtAl2020OneDoesNotSimplyRSVP}. Our findings suggest theta band synchronization as a potential marker of selective exposure, with significant changes when reading attitude-congruent versus attitude-discrepant content. This work establishes a foundation for real-time, multimodal metrics to study selective exposure in dynamic information environments.

    

\begin{figure*}
    \centering
    \includegraphics[width=1.0\linewidth]{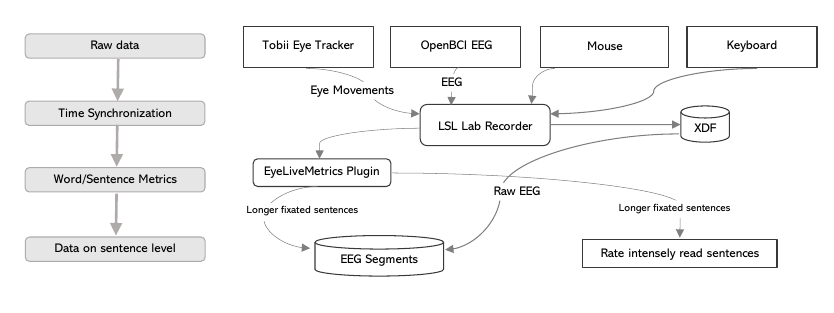}
    \caption{Process to collect different input streams to time-synchronized data and to compute sentence fixation times for the segmentation of EEG data and data annotation.}
    \label{fig:pipeline_technical_setup}
    \Description{TobiiEyeTracker, OpenBCI, Mouse, and Keyboard input streams are broadcast as raw data into LSL. The LSL Lab Recorder application stores the data in a single XDF file per participant and session and also forwards the gaze data to the EyeLiveMetrics browser plugin using WebSocket. EyeLiveMetrics backend extracts saccade and fixation metrics from the raw gaze data. It calculates the longest-fixated sentences, which are presented again to the participants to collect feedback. Further, the word fixation start and end timestamps are used to segment the EEG data from the XDF file.}
\end{figure*}

\section{Metrics for Sensing Selective Exposure}


Based on related work, we propose candidate metrics based on EEG and eye-tracking data to detect selective exposure to capture cognitive and attentional shifts when individuals process congruent versus incongruent news content. Fixation duration indicates attentional bias, where longer fixations on attitude-consistent content and shorter fixations on opposing viewpoints suggest selective exposure~\cite{schmuck2020avoiding}. Theta band power modulation, particularly in the 5–8 Hz range around word onset, reflects differences in cognitive engagement depending on information congruence~\cite{Fischer2013}.

Additionally, we propose investigating Event-Related Potentials (ERPs)~\cite{luck2014introduction} as a measure for selective exposure. ERPs are potentials that correlate with perceived stimuli, including written, spoken, and signed words, drawings, photos, and videos of faces and objects~\cite{annurev:/content/journals/10.1146/annurev.psych.093008.131123}. The cloze probability \cite{hofmann2022language} and the N400 response \cite{Steele2012} provide insights into cognitive effort when processing unexpected words, with higher N400 amplitudes suggesting greater difficulty in integrating incongruent information. Extreme fixation durations, whether longer or shorter than expected, may further indicate points of intervention.

\section{Proposed Experiment}
To validate these metrics, we design an experimental study that integrates EEG and eye-tracking data to examine selective exposure in online news reading.

\subsection{Study Design}
The study will involve 40 to 60 participants with diverse ideological backgrounds. Stimuli will consist of a mix of news articles selected from a large corpus, allowing for natural reading. Instead of relying solely on word fixation duration, we apply a shifting mask through all read sentences to predict next-word probability. Data collection will include continuous EEG recording using a 16-channel cap and real-time eye-tracking to capture reading patterns and cognitive engagement.

\subsection{Apparatus}

We will integrate eye tracking, EEG, mouse, and keyboard inputs into a time-synchronized data stream. \autoref{fig:pipeline_technical_setup} illustrates the data collection workflow. Eye-tracking fixation data will segment EEG recordings, enabling analysis of neural activity associated with specific words and sentences. This multimodal approach links behavioral (gaze), cortical (EEG), and subjective (feedback) data for a comprehensive study of selective exposure mechanisms.

A Tobii Spectrum eye tracker\footnote{\url{https://www.tobii.com/products/eye-trackers/screen-based/tobii-pro-spectrum}} will capture gaze at 300 Hz. EEG will be recorded using semi-dry electrodes with NaCl solution and an OpenBCI Cython board with a Daisy extension\footnote{\url{https://shop.openbci.com/products/all-in-one-gelfree-electrode-cap-bundle}}, providing 16 channels at 125 Hz. Electrodes will be positioned according to the 10-20 system \cite{Jasper1958}, ensuring impedance remains below 20 k$\Omega$. Mouse and keyboard input will be logged at 1000 Hz.

All signals will be streamed via LabStreamingLayer \cite{Kothe2024} (LSL Lab Recorder). Gaze data will be streamed using the Tobii Pro Connector App\footnote{\url{https://github.com/labstreaminglayer/App-TobiiPro}}, EEG via OpenBCI GUI\footnote{\url{https://github.com/OpenBCI/OpenBCI_GUI}}, and keyboard/mouse data via the Input app\footnote{\url{https://github.com/labstreaminglayer/App-Input}}. LabRecorder\footnote{\url{https://github.com/labstreaminglayer/App-LabRecorder}} will combine all streams into a single XDF file. Data alignment will be performed using the python pyxdf package\footnote{\url{https://github.com/xdf-modules/pyxdf}}\footnote{\href{https://labstreaminglayer.readthedocs.io/info/time_synchronization.html}{Time synchronization documentation}}.

Next, we will compute fixation times for words and sentences using EyeLiveMetrics \cite{Hienert2024}\footnote{\url{https://git.gesis.org/iir/eyelivemetrics}}. A fork of the eye-tracking LSL stream will be sent via a web socket to the EyeLiveMetrics browser plugin, which classifies fixations and saccades and maps them to words as Areas of Interest (AOIs). Metrics such as fixation duration and start time will be stored in a database. Sentence-level gaze metrics will be computed, linking eye tracking with EEG-measured cognitive processing and participant ratings.

\subsection{Data Analysis}
EEG preprocessing will involve artifact removal, frequency-domain analysis~\cite{Fischer2013}, and extraction of ERPs~\cite{luck2014introduction}. Eye-tracking data will be analyzed through fixation duration, saccade patterns, and gaze shifts~\cite{schmuck2020avoiding}. We will apply regression models and machine learning classifiers to map EEG and eye-tracking markers to selective exposure tendencies. Furthermore, we will conduct frequentist and Bayesian analysis to evaluate the metrics regarding their prediction reliability of selective exposure. Finally, we will use our results to implement and evaluate future bias-aware systems~\cite{Dingler2023}.



\subsection{Challenges and Considerations}
One of the primary challenges in using EEG and eye tracking for detecting selective exposure is the variability in physiological signals. EEG signals are susceptible to noise from muscle activity, environmental interference, and individual differences in neural processing, limiting their application in real-world applications. In contrast, eye tracking is a less susceptible measurement modality. Previous research examined the correlations between EEG and eye-tracking features~\cite{schneegassEtAl2020BrainCoDe} to determine whether eye-tracking metrics alone are sufficient as a standalone measure. Following this approach, we will apply the same principle in our study on selective exposure. This finding can establish eye tracking as a single measure for selective exposure in the real world, where EEG is used for experimental settings. 

Furthermore, the use of physiological sensing in online news consumption raises ethical concerns regarding user privacy and consent. EEG and eye-tracking data can reveal political preferences that are sensitive information for individuals that can lead to privacy violations whose scope can be hardly understood by users (cf.~\cite{windl2025illusion, schaub2016watching}). Consequently, we will conduct literature research concerning the handling of privacy when using physiological sensing to gain insights into user states. 

\section{Conclusion}
This paper presents an experiment utilizing electroencephalography and eye tracking to assess participants' susceptibility to selective exposure, a cognitive bias that leads individuals to prefer information that aligns with their existing beliefs while disregarding opposing perspectives. We consider gaze fixations, saccades, theta power, and event-related potentials as metrics for sensing selective exposure when reading online news. The final goal of this research project is to isolate suitable features that predict the susceptibility of selective exposure. These features can be used in future bias-aware systems to increase users sensitivity for unbalanced information diets.

\bibliographystyle{ACM-Reference-Format}
\bibliography{main}


\begin{thebibliography}{22}


\ifx \showCODEN    \undefined \def \showCODEN     #1{\unskip}     \fi
\ifx \showISBNx    \undefined \def \showISBNx     #1{\unskip}     \fi
\ifx \showISBNxiii \undefined \def \showISBNxiii  #1{\unskip}     \fi
\ifx \showISSN     \undefined \def \showISSN      #1{\unskip}     \fi
\ifx \showLCCN     \undefined \def \showLCCN      #1{\unskip}     \fi
\ifx \shownote     \undefined \def \shownote      #1{#1}          \fi
\ifx \showarticletitle \undefined \def \showarticletitle #1{#1}   \fi
\ifx \showURL      \undefined \def \showURL       {\relax}        \fi
\providecommand\bibfield[2]{#2}
\providecommand\bibinfo[2]{#2}
\providecommand\natexlab[1]{#1}
\providecommand\showeprint[2][]{arXiv:#2}

\bibitem[Boonprakong et~al\mbox{.}(2023)]%
        {Dingler2023}
\bibfield{author}{\bibinfo{person}{Nattapat Boonprakong}, \bibinfo{person}{Xiuge Chen}, \bibinfo{person}{Catherine Davey}, \bibinfo{person}{Benjamin Tag}, {and} \bibinfo{person}{Tilman Dingler}.} \bibinfo{year}{2023}\natexlab{}.
\newblock \showarticletitle{Bias-Aware Systems: Exploring Indicators for the Occurrences of Cognitive Biases When Facing Different Opinions}. In \bibinfo{booktitle}{\emph{Proceedings of the 2023 CHI Conference on Human Factors in Computing Systems}}. \bibinfo{pages}{1--19}.
\newblock


\bibitem[{Boonprakong, Nattapat and Pareek, Saumya and Tag, Benjamin and Goncalves, Jorge and Dingler, Tilman}(2025)]%
        {boonprakong2025assessing}
\bibfield{author}{\bibinfo{person}{{Boonprakong, Nattapat and Pareek, Saumya and Tag, Benjamin and Goncalves, Jorge and Dingler, Tilman}}.} \bibinfo{year}{2025}\natexlab{}.
\newblock \showarticletitle{{Assessing Susceptibility Factors of Confirmation Bias in News Feed Reading}}. In \bibinfo{booktitle}{\emph{{Proceedings of the 2025 CHI Conference on Human Factors in Computing Systems}}}. \bibinfo{publisher}{ACM}, \bibinfo{address}{New York, NY, USA}, \bibinfo{numpages}{19}~pages.
\newblock
\showISBNx{979-8-4007-1394-1/25/04}
\href{https://doi.org/10.1145/3706598.3713873}{doi:\nolinkurl{10.1145/3706598.3713873}}


\bibitem[{Boonprakong, Nattapat and Tag, Benjamin and Goncalves, Jorge and Dingler, Tilman}(2025)]%
        {boonprakong4797710scopingreview}
\bibfield{author}{\bibinfo{person}{{Boonprakong, Nattapat and Tag, Benjamin and Goncalves, Jorge and Dingler, Tilman}}.} \bibinfo{year}{2025}\natexlab{}.
\newblock \showarticletitle{{How Do HCI Researchers Study Cognitive Biases? A Scoping Review}}. In \bibinfo{booktitle}{\emph{{Proceedings of the 2025 CHI Conference on Human Factors in Computing Systems}}}. \bibinfo{publisher}{ACM}, \bibinfo{address}{New York, NY, USA}, \bibinfo{numpages}{20}~pages.
\newblock
\showISBNx{979-8-4007-1394-1/25/04}
\href{https://doi.org/10.1145/3706598.3713450}{doi:\nolinkurl{10.1145/3706598.3713450}}


\bibitem[Carroll and Olson(1988)]%
        {CARROLL198845}
\bibfield{author}{\bibinfo{person}{John~M. Carroll} {and} \bibinfo{person}{Judith~Reitman Olson}.} \bibinfo{year}{1988}\natexlab{}.
\newblock \showarticletitle{Chapter 2 - Mental Models in Human-Computer Interaction11This chapter appeared in its entirety and is reprinted from Mental Models in Human Computer Interaction: Research Issues about What the User of Software Knows, J.M. Carroll and J.R. Olson, Editors,-The report of the workshop on software human factors: Users mental models, Nancy Anderson, chair, sponsored by the Committee on Human Factors, Commission on Behavioral and Social Sciences and Education, National Research Council, published by the National Academy Press, 1987}.
\newblock In \bibinfo{booktitle}{\emph{Handbook of Human-Computer Interaction}}, \bibfield{editor}{\bibinfo{person}{MARTIN HELANDER}} (Ed.). \bibinfo{publisher}{North-Holland}, \bibinfo{address}{Amsterdam}, \bibinfo{pages}{45--65}.
\newblock
\showISBNx{978-0-444-70536-5}
\href{https://doi.org/10.1016/B978-0-444-70536-5.50007-5}{doi:\nolinkurl{10.1016/B978-0-444-70536-5.50007-5}}


\bibitem[Chiossi et~al\mbox{.}(2025)]%
        {chiossi2025designing}
\bibfield{author}{\bibinfo{person}{Francesco Chiossi}, \bibinfo{person}{Changkun Ou}, \bibinfo{person}{Carolina Gerhardt}, \bibinfo{person}{Felix Putze}, {and} \bibinfo{person}{Sven Mayer}.} \bibinfo{year}{2025}\natexlab{}.
\newblock \showarticletitle{Designing and Evaluating an Adaptive Virtual Reality System using EEG Frequencies to Balance Internal and External Attention States}.
\newblock \bibinfo{journal}{\emph{International Journal of Human-Computer Studies}}  \bibinfo{volume}{196} (\bibinfo{year}{2025}), \bibinfo{pages}{103433}.
\newblock
\href{https://doi.org/10.1016/j.ijhcs.2024.103433}{doi:\nolinkurl{10.1016/j.ijhcs.2024.103433}}


\bibitem[Fischer et~al\mbox{.}(2013)]%
        {Fischer2013}
\bibfield{author}{\bibinfo{person}{Peter Fischer}, \bibinfo{person}{Matthias Reinweber}, \bibinfo{person}{Claudia Vogrincic}, \bibinfo{person}{Axel Schäfer}, \bibinfo{person}{Anne Schienle}, {and} \bibinfo{person}{Gregor Volberg}.} \bibinfo{year}{2013}\natexlab{}.
\newblock \showarticletitle{Neural mechanisms of selective exposure: An EEG study on the processing of decision-consistent and inconsistent information}.
\newblock \bibinfo{journal}{\emph{International Journal of Psychophysiology}} \bibinfo{volume}{87}, \bibinfo{number}{1} (\bibinfo{year}{2013}), \bibinfo{pages}{13--18}.
\newblock
\showISSN{0167-8760}
\href{https://doi.org/10.1016/j.ijpsycho.2012.10.011}{doi:\nolinkurl{10.1016/j.ijpsycho.2012.10.011}}


\bibitem[Foulsham et~al\mbox{.}(2013)]%
        {foulsham2013mind}
\bibfield{author}{\bibinfo{person}{Tom Foulsham}, \bibinfo{person}{James Farley}, {and} \bibinfo{person}{Alan Kingstone}.} \bibinfo{year}{2013}\natexlab{}.
\newblock \showarticletitle{Mind wandering in sentence reading: decoupling the link between mind and eye.}
\newblock \bibinfo{journal}{\emph{Canadian Journal of Experimental Psychology/Revue canadienne de psychologie exp{\'e}rimentale}} \bibinfo{volume}{67}, \bibinfo{number}{1} (\bibinfo{year}{2013}), \bibinfo{pages}{51}.
\newblock


\bibitem[Hienert et~al\mbox{.}(2024)]%
        {Hienert2024}
\bibfield{author}{\bibinfo{person}{Daniel Hienert}, \bibinfo{person}{Heiko Schmidt}, \bibinfo{person}{Thomas Kr\"{a}mer}, {and} \bibinfo{person}{Dagmar Kern}.} \bibinfo{year}{2024}\natexlab{}.
\newblock \showarticletitle{EyeLiveMetrics: Real-time Analysis of Online Reading with Eye Tracking}. In \bibinfo{booktitle}{\emph{Proceedings of the 2024 Symposium on Eye Tracking Research and Applications}} (Glasgow, United Kingdom) \emph{(\bibinfo{series}{ETRA '24})}. \bibinfo{publisher}{Association for Computing Machinery}, \bibinfo{address}{New York, NY, USA}, Article \bibinfo{articleno}{81}, \bibinfo{numpages}{7}~pages.
\newblock
\showISBNx{9798400706073}
\href{https://doi.org/10.1145/3649902.3656495}{doi:\nolinkurl{10.1145/3649902.3656495}}


\bibitem[Hofmann et~al\mbox{.}(2022)]%
        {hofmann2022language}
\bibfield{author}{\bibinfo{person}{Markus~J Hofmann}, \bibinfo{person}{Steffen Remus}, \bibinfo{person}{Chris Biemann}, \bibinfo{person}{Ralph Radach}, {and} \bibinfo{person}{Lars Kuchinke}.} \bibinfo{year}{2022}\natexlab{}.
\newblock \showarticletitle{Language models explain word reading times better than empirical predictability}.
\newblock \bibinfo{journal}{\emph{Frontiers in Artificial Intelligence}}  \bibinfo{volume}{4} (\bibinfo{year}{2022}), \bibinfo{pages}{730570}.
\newblock


\bibitem[Jasper(1958)]%
        {Jasper1958}
\bibfield{author}{\bibinfo{person}{Herbert Jasper}.} \bibinfo{year}{1958}\natexlab{}.
\newblock \showarticletitle{Report of the committee on methods of clinical examination in electroencephalography: 1957}.
\newblock \bibinfo{journal}{\emph{Electroencephalography and Clinical Neurophysiology}} \bibinfo{volume}{10}, \bibinfo{number}{2} (\bibinfo{year}{1958}), \bibinfo{pages}{370--375}.
\newblock
\showISSN{0013-4694}
\href{https://doi.org/10.1016/0013-4694(58)90053-1}{doi:\nolinkurl{10.1016/0013-4694(58)90053-1}}


\bibitem[Klimesch et~al\mbox{.}(2001)]%
        {Klimesch2001}
\bibfield{author}{\bibinfo{person}{W. Klimesch}, \bibinfo{person}{M. Doppelmayr}, \bibinfo{person}{W. Stadler}, \bibinfo{person}{D. Pöllhuber}, \bibinfo{person}{P. Sauseng}, {and} \bibinfo{person}{D. Röhm}.} \bibinfo{year}{2001}\natexlab{}.
\newblock \showarticletitle{Episodic retrieval is reflected by a process specific increase in human electroencephalographic theta activity}.
\newblock \bibinfo{journal}{\emph{Neuroscience Letters}} \bibinfo{volume}{302}, \bibinfo{number}{1} (\bibinfo{year}{2001}), \bibinfo{pages}{49--52}.
\newblock
\showISSN{0304-3940}
\href{https://doi.org/10.1016/S0304-3940(01)01656-1}{doi:\nolinkurl{10.1016/S0304-3940(01)01656-1}}


\bibitem[Kosch et~al\mbox{.}(2020)]%
        {KoschEtAl2020OneDoesNotSimplyRSVP}
\bibfield{author}{\bibinfo{person}{Thomas Kosch}, \bibinfo{person}{Albrecht Schmidt}, \bibinfo{person}{Simon Thanheiser}, {and} \bibinfo{person}{Lewis~L. Chuang}.} \bibinfo{year}{2020}\natexlab{}.
\newblock \showarticletitle{One does not Simply RSVP: Mental Workload to Select Speed Reading Parameters using Electroencephalography}. In \bibinfo{booktitle}{\emph{Proceedings of the 2020 CHI Conference on Human Factors in Computing Systems}} (Honolulu, HI, USA) \emph{(\bibinfo{series}{CHI '20})}. \bibinfo{publisher}{Association for Computing Machinery}, \bibinfo{address}{New York, NY, USA}, \bibinfo{pages}{1–13}.
\newblock
\showISBNx{9781450367080}
\href{https://doi.org/10.1145/3313831.3376766}{doi:\nolinkurl{10.1145/3313831.3376766}}


\bibitem[Kothe et~al\mbox{.}(2024)]%
        {Kothe2024}
\bibfield{author}{\bibinfo{person}{Christian Kothe}, \bibinfo{person}{Seyed~Yahya Shirazi}, \bibinfo{person}{Tristan Stenner}, \bibinfo{person}{David Medine}, \bibinfo{person}{Chadwick Boulay}, \bibinfo{person}{Matthew~I. Grivich}, \bibinfo{person}{Tim Mullen}, \bibinfo{person}{Arnaud Delorme}, {and} \bibinfo{person}{Scott Makeig}.} \bibinfo{year}{2024}\natexlab{}.
\newblock \bibinfo{title}{The Lab Streaming Layer for Synchronized Multimodal Recording}.
\newblock
\href{https://doi.org/10.1101/2024.02.13.580071}{doi:\nolinkurl{10.1101/2024.02.13.580071}}
\showeprint{https://www.biorxiv.org/content/early/2024/02/14/2024.02.13.580071.full.pdf}


\bibitem[Krämer et~al\mbox{.}(2025)]%
        {kraemer2025escaping}
\bibfield{author}{\bibinfo{person}{Thomas Krämer}, \bibinfo{person}{Daniel Hienert}, \bibinfo{person}{Francesco Chiossi}, \bibinfo{person}{Thomas Kosch}, {and} \bibinfo{person}{Dagmar Kern}.} \bibinfo{year}{2025}\natexlab{}.
\newblock \showarticletitle{{Escaping the Filter Bubble: Evaluating Electroencephalographic Theta Band Synchronization as Indicator for Selective Exposure in Online News Reading}}. In \bibinfo{booktitle}{\emph{{Extended Abstracts of the 2025 CHI Conference on Human Factors in Computing Systems}}} \emph{(\bibinfo{series}{CHI EA '25})}. \bibinfo{publisher}{ACM}, \bibinfo{address}{New York, NY, USA}.
\newblock
\showISBNx{979-8-4007-1394-1/25/04}
\href{https://doi.org/10.1145/3706599.3720078}{doi:\nolinkurl{10.1145/3706599.3720078}}


\bibitem[Kutas and Federmeier(2011)]%
        {annurev:/content/journals/10.1146/annurev.psych.093008.131123}
\bibfield{author}{\bibinfo{person}{Marta Kutas} {and} \bibinfo{person}{Kara~D. Federmeier}.} \bibinfo{year}{2011}\natexlab{}.
\newblock \showarticletitle{Thirty Years and Counting: Finding Meaning in the N400 Component of the Event-Related Brain Potential (ERP)}.
\newblock \bibinfo{journal}{\emph{Annual Review of Psychology}} \bibinfo{volume}{62}, \bibinfo{number}{Volume 62, 2011} (\bibinfo{year}{2011}), \bibinfo{pages}{621--647}.
\newblock
\showISSN{1545-2085}
\href{https://doi.org/10.1146/annurev.psych.093008.131123}{doi:\nolinkurl{10.1146/annurev.psych.093008.131123}}


\bibitem[Long et~al\mbox{.}(2024)]%
        {long2024multimodal}
\bibfield{author}{\bibinfo{person}{Xingyu Long}, \bibinfo{person}{Sven Mayer}, {and} \bibinfo{person}{Francesco Chiossi}.} \bibinfo{year}{2024}\natexlab{}.
\newblock \showarticletitle{Multimodal Detection of External and Internal Attention in Virtual Reality using EEG and Eye Tracking Features}. In \bibinfo{booktitle}{\emph{Proceedings of Mensch Und Computer 2024}} (Karlsruhe, Germany) \emph{(\bibinfo{series}{MuC '24})}. \bibinfo{publisher}{Association for Computing Machinery}, \bibinfo{address}{New York, NY, USA}, \bibinfo{pages}{29–43}.
\newblock
\showISBNx{9798400709982}
\href{https://doi.org/10.1145/3670653.3670657}{doi:\nolinkurl{10.1145/3670653.3670657}}


\bibitem[Luck(2014)]%
        {luck2014introduction}
\bibfield{author}{\bibinfo{person}{Steven~J Luck}.} \bibinfo{year}{2014}\natexlab{}.
\newblock \bibinfo{booktitle}{\emph{An introduction to the event-related potential technique}}.
\newblock \bibinfo{publisher}{MIT press}.
\newblock


\bibitem[Schaub et~al\mbox{.}(2016)]%
        {schaub2016watching}
\bibfield{author}{\bibinfo{person}{Florian Schaub}, \bibinfo{person}{Aditya Marella}, \bibinfo{person}{Pranshu Kalvani}, \bibinfo{person}{Blase Ur}, \bibinfo{person}{Chao Pan}, \bibinfo{person}{Emily Forney}, {and} \bibinfo{person}{Lorrie~Faith Cranor}.} \bibinfo{year}{2016}\natexlab{}.
\newblock \showarticletitle{Watching them watching me: Browser extensions’ impact on user privacy awareness and concern}. In \bibinfo{booktitle}{\emph{NDSS workshop on usable security}}, Vol.~\bibinfo{volume}{10}.
\newblock
\href{https://doi.org/10.14722/usec.2016.23017}{doi:\nolinkurl{10.14722/usec.2016.23017}}


\bibitem[Schmuck et~al\mbox{.}(2020)]%
        {schmuck2020avoiding}
\bibfield{author}{\bibinfo{person}{Desir{\'e}e Schmuck}, \bibinfo{person}{Miriam Tribastone}, \bibinfo{person}{J{\"o}rg Matthes}, \bibinfo{person}{Franziska Marquart}, {and} \bibinfo{person}{Eva~Maria Bergel}.} \bibinfo{year}{2020}\natexlab{}.
\newblock \showarticletitle{Avoiding the other side? An eye-tracking study of selective exposure and selective avoidance effects in response to political advertising.}
\newblock \bibinfo{journal}{\emph{Journal of Media Psychology: Theories, Methods, and Applications}} \bibinfo{volume}{32}, \bibinfo{number}{3} (\bibinfo{year}{2020}), \bibinfo{pages}{158}.
\newblock


\bibitem[Schneegass et~al\mbox{.}(2020)]%
        {schneegassEtAl2020BrainCoDe}
\bibfield{author}{\bibinfo{person}{Christina Schneegass}, \bibinfo{person}{Thomas Kosch}, \bibinfo{person}{Andrea Baumann}, \bibinfo{person}{Marius Rusu}, \bibinfo{person}{Mariam Hassib}, {and} \bibinfo{person}{Heinrich Hussmann}.} \bibinfo{year}{2020}\natexlab{}.
\newblock \showarticletitle{BrainCoDe: Electroencephalography-based Comprehension Detection during Reading and Listening}. In \bibinfo{booktitle}{\emph{Proceedings of the 2020 CHI Conference on Human Factors in Computing Systems}} (Honolulu, HI, USA) \emph{(\bibinfo{series}{CHI '20})}. \bibinfo{publisher}{Association for Computing Machinery}, \bibinfo{address}{New York, NY, USA}, \bibinfo{pages}{1–13}.
\newblock
\showISBNx{9781450367080}
\href{https://doi.org/10.1145/3313831.3376707}{doi:\nolinkurl{10.1145/3313831.3376707}}


\bibitem[Steele et~al\mbox{.}(2012)]%
        {Steele2012}
\bibfield{author}{\bibinfo{person}{Vaughn~R. Steele}, \bibinfo{person}{Edward~M. Bernat}, \bibinfo{person}{Paul van~den Broek}, \bibinfo{person}{Paul~F. Collins}, \bibinfo{person}{Christopher~J. Patrick}, {and} \bibinfo{person}{Chad~J. Marsolek}.} \bibinfo{year}{2012}\natexlab{}.
\newblock \showarticletitle{Separable processes before, during, and after the N400 elicited by previously inferred and new information: Evidence from time-frequency decompositions}.
\newblock \bibinfo{journal}{\emph{Brain Research}}  \bibinfo{volume}{1492} (\bibinfo{year}{2012}), \bibinfo{pages}{92--107}.
\newblock
\href{https://doi.org/10.1016/j.brainres.2012.11.016}{doi:\nolinkurl{10.1016/j.brainres.2012.11.016}}
\newblock
\shownote{https://europepmc.org/articles/pmc3534777?pdf=render}.


\bibitem[Windl et~al\mbox{.}(2025)]%
        {windl2025illusion}
\bibfield{author}{\bibinfo{person}{Maximiliane Windl}, \bibinfo{person}{Roman Amberg}, {and} \bibinfo{person}{Thomas Kosch}.} \bibinfo{year}{2025}\natexlab{}.
\newblock \showarticletitle{{The Illusion of Privacy: Investigating User Misperceptions in Browser Tracking Protection}}. In \bibinfo{booktitle}{\emph{{Proceedings of the 2025 CHI Conference on Human Factors in Computing Systems}}}. \bibinfo{publisher}{ACM}, \bibinfo{address}{New York, NY, USA}.
\newblock
\showISBNx{979-8-4007-1394-1/25/04}
\href{https://doi.org/10.1145/3706598.3713912}{doi:\nolinkurl{10.1145/3706598.3713912}}


\end{thebibliography}
\end{document}